# Significance of Mobility on Received Signal Strength: An Experimental Investigation

Pavan Kumar Pedapolu, Pradeep Kumar, Vaidya Harish, Satvik Venturi, Sushil Kumar Bharti[#],
Vinay Kumar, and Sudhir Kumar
[#]Department of Electrical Engineering, College of Engineering and Management, Kolaghat
Department of Electronics and Communication Engineering, Visvesvaraya National Institute of Technology, Nagpur, India
Email: {pavan.pedapolu, pradeepvnit13, vaidyaharish849, satvikventuri, sushilkumarbharti2, vinayrel01, sudhirk.iitk}@gmail.com

*Abstract* - **In this paper, estimation of mobility using received signal strength is presented. In contrast to standard methods, speed can be inferred without the use of any additional hardware like accelerometer, gyroscope or position estimator. The strength of Wi-Fi signal is considered herein to compute the time-domain features such as mean, minimum, maximum, and autocorrelation. The experiments are carried out in different environments like academic area, residential area and in open space. The complexity of the algorithm in training and testing phase are quadratic and linear with the number of Wi-Fi samples respectively. The experimental results indicate that the average error in the estimated speed is 12 % when the maximum signal strength features are taken into account. The proposed method is cost-effective and having a low complexity with reasonable accuracy in a Wi-Fi or cellular environment. Additionally, the proposed method is scalable that is the performance is not affected in a multi-smartphones scenario.**

## I. INTRODUCTION:

A large number of applications in today's modern world requires internet connectivity. The connectivity is of two types i.e. wired and wireless. There are many advantages in wireless connection such as mobility and accessibility at any place. Wi-Fi signal is one of the most used wireless connectivity [1-2]. Wi-Fi signal is almost present at all places due to its applications. In this modern life, every person uses a mobile phone which definitely requires a Wi-Fi signal for connectivity. The performance of Wi-Fi signal greatly depends on the signal strength. The signal strength determines the data rate. Wi-Fi signal strength is affected by several factors like clutter, interference of other Wi-Fi signal and ground reflected signal. People and vehicles with Wi-Fi enabled devices move from one place to another so there arises the dependency of Wi-Fi signal strength on speed. Additionally, in the moving vehicles, we observe that signals from the cellular base station vary quite considerably. It leads to the packet drop and the seamless wireless connectivity is not established. The data rate decreases with increase of the speed of the vehicle.

In [3], walking direction is estimated using a device-free method, however, the speed with which it is moving in that direction is not estimated. The detection of vehicle and the estimation of speed is presented in [4]. The speed is determined with signal strength variance using a multi-class SVM based approach. However, the pattern of the variance with increasing or decreasing speed is not reflected in our environmental settings. The channel state information based approach is proposed in [5]. The CSI based methods requires Wi-Fi network interface cards (NIC) or 802.11 a/g/n wireless connection. A correlation based method wherein two time-series signal strength data are matched is described in [6].

In this paper, we compute the features of the signal that has specific pattern with increase of speed. In particular, time-domain features such as maximum, and minimum signal strengths and the coefficient of autocorrelation are computed. The coefficient of autocorrelation also provides the insight of greater and lower speeds. On the other hand, given the speed of the vehicles, the maximum and minimum signal strength within certain time interval is estimated. This provides us the insight about the data rate and hence the wireless connectivity. The Signal-to-Noise Ratio (SNR) may also be taken into account to provide the better estimate of wireless connectivity. The proposed method is suitable for either Wi-Fi or cellular environment. The method has a potential applications in a smart traffic monitoring, 3G/4G/5G networks and pedestrian tracking in an indoor settings. The complexity of the proposed algorithm for estimating the speed is lesser since the equation derived from the data is a linear one.

The rest of the paper is organized as follows: In Section II, time domain features for estimation of mobility is described. A detailed experimental analysis of received Wi-Fi signal strength with mobility is provided. Experimental setup, and experimental results are presented in Section III. The analysis is carried out by capturing the raw data of Wi-Fi signal through mobile phone by varying the speed. A brief conclusion and future work are briefed in Section IV.

## II. METHODOLOGY FOR ESTIMATION OF MOBILITY:

In this section, first the definitions of time domain features [7] are presented. The time domain features are mean, variance, skewness, kurtosis, higher order moments, maximum, minimum and coefficient of autocorrelation as summarized in Table 1. Subsequently, mathematical relationship between features and speed is described. A brief algorithmic description of the proposed algorithm is then followed.

a. **Mean:** It defined as the average value of the Wi-Fi samples collected within certain time interval. It is calculated by taking the sum of all the Wi-Fi data samples divided by total number of samples.

Table 1: Time domain and frequency domain features for extracting mobility pattern

| Features | | Remarks/Provides Pattern? |
|---|---|---|
| *Time domain features*: | Mean | Approximately constant |
| | Variance | No |
| | Skewness | No |
| | Kurtosis | No |
| | Minimum | Yes |
| | Maximum | Yes |
| | Autocorrelation | Yes |
| *Frequency domain feature*: | FFT | No |

b. **Maximum and Minimum Signal Strength:** In any set of signal samples, the highest value among all the samples is considered as the maximum signal strength. Similarly the lowest among all the Wi-Fi samples is considered as the minimum signal strength. The models are established in two different environments. The Least –Square method is utilized for establishing these expressions. It may be noted that the value of the coefficients are very close across the sites.

| For site 1 | For site 2 |
|---|---|
| $\max(S) = a_1 v + b_1$ | $\max(S) = a_3 v + b_3$ |
| $a_1 = -3.334$ | $a_3 = -4.145$ |
| $b_1 = -34.27$ | $b_3 = -32.6$ |
| SSE = 15.28 | SSE = 8.691 |
| R-Squared = 0.853 | R-Squared = 0.9336 |
| Adjusted R-Squared = 0.8163 | Adjusted R-Squared = 0.917 |
| RMSE = 0.955 | RMSE = 1.474 |
| | |
| $\min(S) = a_2 v + b_2$ | $\min(S) = a_4 v + b_4$ |
| $a_2 = 2.614$ | $a_4 = 2.34$ |
| $b_2 = -74.7$ | $b_4 = -77.15$ |
| SSE = 26.96 | SSE = 34.59 |
| R-Squared = 0.6692 | R-Squared = 0.5294 |
| Adjusted R-Squared = 0.5865 | Adjusted R-Squared = 0.4118 |
| RMSE = 2.596 | RMSE = 2.941 |

Where $S$ denotes Wi-Fi signal strength in dBm and $v$ represents the speed in m/s. $a$ and $b$ are constants of the models.

Time-domain features are modelled as a linear plot with speed for extracting a trend. Since it involves approximation, it is necessary to know the extent of correctness of those models which is justified by the four mentioned Goodness of fit statistics which are as follows:

**Sum Squared Error:** It is the measure of deviation of the response from the fit. A value closer to 0 indicates that the model has a smaller random error component, and that the fit will be more useful for prediction.

**R-Squared:** It measures how successfully the fit explaining the variation of the data, it is achieved by squaring the correlation between the response values and the predicted response values. Its value lies between 0 and 1, as the value goes closer to 1 means greater proportion of variance is counted and is desirable.

**Adjusted R-Squared:** This statistic uses the above defined R-Squared value, and adjusts it based on the residual degrees of freedom. The residual degrees of freedom is defined as the difference between number of response values and the number of fitted coefficients estimated from the response values.

**Root Mean Square Error:** It is an estimate of the standard deviation of random component in the data.

**Coefficient of Autocorrelation:**
Correlation is a statistical technique that depicts the similarity between two signals. Hence autocorrelation gives the similarity between the present signal samples and its time delayed samples. Correlation coefficient gives the interdependence of two set of data samples. The correlation coefficient decreases exponential as

$$\rho = ax^b + c$$

Where $\rho$ and x denote the correlation coefficient and lag respectively. *a, b, c* are the coefficient of the model.

Table 2. Values of the parameters *a, b, c.* $v_1, v_2, v_3, v_4, v_5$ are the speeds that are considered for site 1 and site 2.

|  | Site 1 | | | Site 2 | | |
| --- | --- | --- | --- | --- | --- | --- |
|  | a | B | c | a | b | c |
| $v_1$ | -0.003006 | 1.292 | 1.002 | -0.002549 | 1.285 | 1.002 |
| $v_2$ | -0.004192 | 1.403 | 1 | -0.004411 | 1.263 | 1.003 |
| $v_3$ | -0.005066 | 1.392 | 1.001 | -0.005858 | 1.33 | 1.002 |
| $v_4$ | -0.0112 | 1.269 | 1.007 | -0.01081 | 1.188 | 1.012 |
| $v_5$ | -0.01476 | 1.257 | 1.01 | -0.01546 | 1.213 | 1.009 |

The algorithm for computation of features and subsequently speed is summarized as follows:

*Algorithm 1: Speed Estimation using Time-Domain Features of Wi-Fi Signals*

1. **INITIALISATION:** Mobile phone with samples collecting software application and access point which provides the Wi-Fi signal are taken.
2. **DATA COLLECTION:** The Wi-Fi samples are taken from the mobile phone for different sites, speeds are measured and raw data is modelled to compute the time domain features.
3. **COMPUTATIONAL FEATURES:** Maximum and minimum signal strength, autocorrelation are computed.
4. **ESTIMATION:** Time domain features like maximum signal strength, minimum signal strengths and the correlation coefficients are utilized to compute the speed.
5. **OUTPUT:** Speed of the receiver.

It may be noted that the mean is found to be constant with varying speed. The variance, skewness, kurtosis, higher order moments, and frequency domain features do not give any insight of the mobility.

## III. PERFORMANCE EVALUATION:

In this section, real experimental setup and experimental results for mobility estimation using wifi signal is described in the ensuing section.

**1. EXPERIMENTAL SETUP:**
There are various apps like *wifi analyser app, wifi data set app* are taken into account to get Wi-Fi samples. However, these apps could not provide required number of samples for analysis. Therefore, the Wi-Fi samples are collected by using an app *SensorAp*p developed by Prof. Niki Trigoni and her research group at Oxford University, United Kingdom. The device used for receiving the Wi-Fi signal strength is Le X526.The source of Wi-Fi signal used is hosted network created in Lenovo S-50 as shown in Fig. 1.

In order to get various speeds, a robot is designed such that it moves with different speeds and holds mobile phone on it. Additionally, the readings are taken manually moving with different speeds in different direction towards the source kept at a distance as shown in the Fig. 1. The experiments are conducted using real setup at Department of Electronics and Communication Engineering at Visvesvaraya National Institute of Technology, Nagpur. A person holding the mobile phone in his hand covered a distance of 30 m with five different speeds and the source is placed in the middle of the total distance covered. Many trials are taken to show the effectiveness of the proposed algorithm. The different speeds for which the samples are collected are 0.6 m/s, 1 m/s, 1.42 m/s, 2 m/s, 3 m/s and 3.75 m/s. It may be noted that, speed in order of Kmph using fast moving vehicles may be considered. In that scenario, cellular base station may be taken as the transmitter, while, the mobile phone is considered as the receiver.

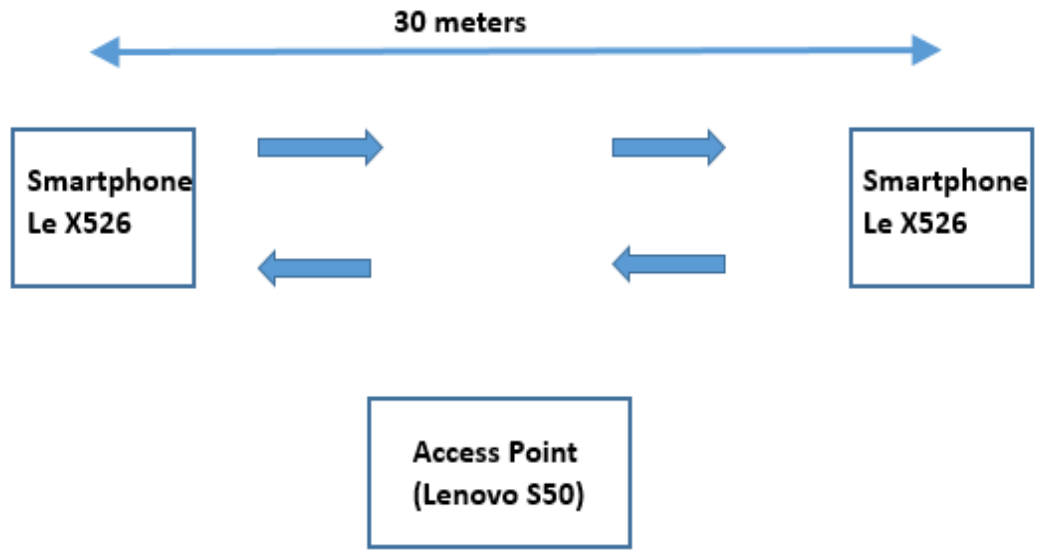

Figure 1: *Experimental setup consisting of Access point (laptop) and mobile phones for collection of Wi-Fi samples at different speeds.*

## 2. EXPERIMENTAL RESULTS:

The features considered in this work are time domain features of Wi-Fi signal with different speeds. Wi-Fi can be analysed through its signal strength. It is fascinating to know how Wi-Fi signal strength varies with different speeds.

### A. Time domain features:

**Mean:** Since we are working with the large set of Wi-Fi signal samples, mean is the best way to represent this data samples with single value. The experiments that are conducted shows that mean of different set of samples is nearly constant with various speeds considered, as expected. This is illustrated from the results obtained in two sites as shown in Fig 2.

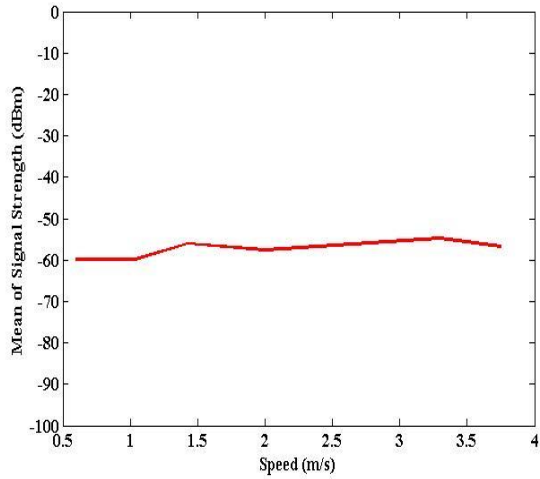 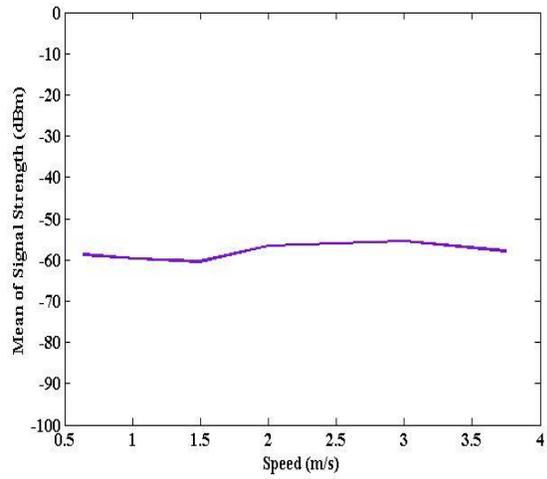

*Fig 2.1: Mean vs speed plot (site 1)*  *Fig 2.2: Mean vs Speed plot (site 2)*

**Maximum and Minimum Signal Strength:**

In order to have a good idea about the data rate, it is necessary to know about the maximum and minimum signal strength. As different speeds are taken into account, maximum and minimum signal strength play an important role in improving the range of the signal as shown in Fig. 3 and Fig. 4.

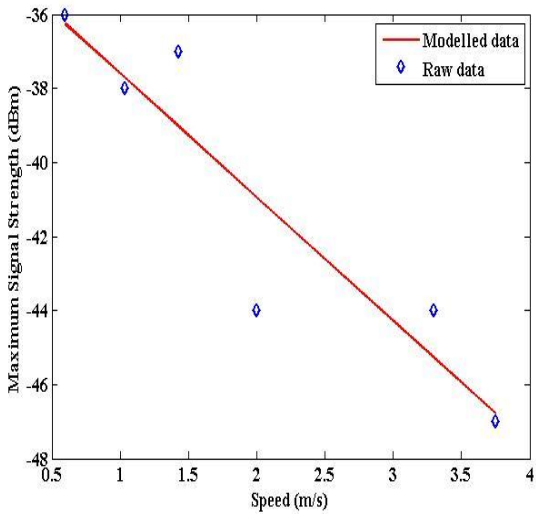 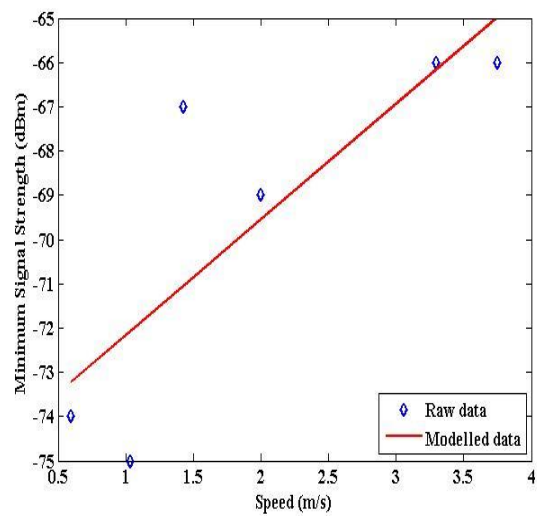

*Fig 3: Variation of maximum & minimum signal strengths with speed for site 1*

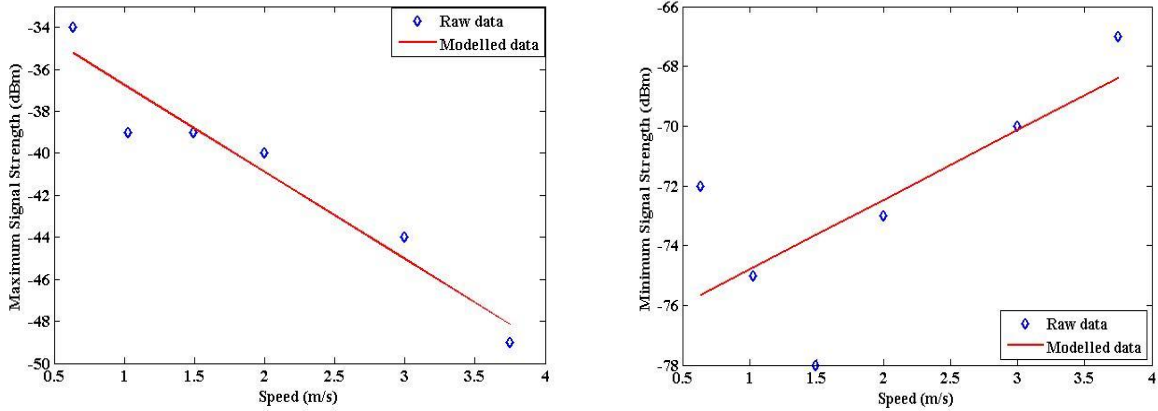

*Fig 4: Variation of maximum & minimum signal strengths with speed for site 2*

When the speed of Wi-Fi signal receiver increases maximum signal strength decreases and minimum signal strength increases. The above results depict that when a person is moving slowly, signal strength will be maximum hence data rate is high and vice versa. Mathematical analysis of the above data shows that the signal strength varies linearly with different range of speeds which is described in Section II.

**Autocorrelation:**
It is used to measure the similarities between present samples and past samples obtained in Fig. 5 It provides the insight about the degree of randomness of the signal. If the signal is not random and follows a pattern then it can be represented mathematically as expressed in Section II and its features can be analysed.

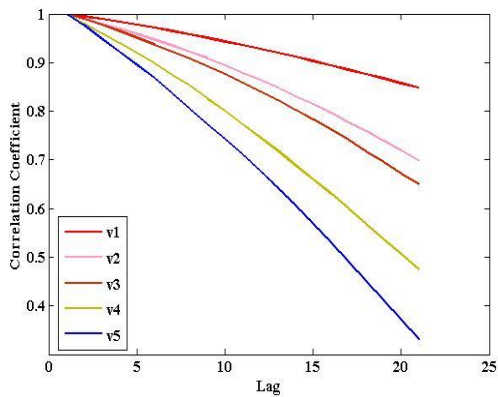
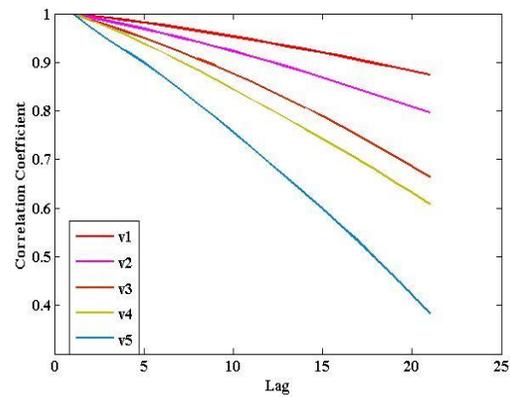

*Fig 5.1: Autocorrelation for site 1*          *Fig 5.2: Autocorrelation for site 2*

From the above graph we can conclude that with increase in speed ($v_1 < v_2 < v_3 < v_4 < v_5$), randomness in Wi-Fi signal also increases.

**Testing Phase:**
We calculate the speed from the equations of maximum and minimum signal strengths time domain features. Considering the maximum signal strength feature equations for two sites, the speed can be expressed as

**Site: 1**
$$v = \frac{\max(s) - b_1}{a_1}$$
Where $a_1 = -3.334$
$b_1 = -34.27$

**Site: 2**
$$v = \frac{\max(s) - b_3}{a_3}$$
Where $a_3 = -4.145$
$b_3 = -32.6$

Considering the minimum signal strength feature equations for two sites, the speed can be estimated by

**Site: 1**
$$v = \frac{\min(s) - b_2}{a_2}$$
Where $a_2 = 2.614$
$b_2 = -74.7$

**Site: 2**
$$v = \frac{\min(s) - b_4}{a_4}$$
Where $a_4 = 2.34$
$b_4 = -77.15$

Root Mean Square Error (RMSE) for maximum signal strength are 0.473 and 0.445 for site 1 and site 2 respectively. Similarly, RMSE for minimum signal strength are 1.516 and 1.549 for site 1 and site 2 respectively. RMSE of the speed from minimum signal strength equation is greater than that of the maximum signal strength because the noise is more when the receiver move away from the access point. Hence, the maximum signal strength time domain feature outperforms the minimum signal strength time domain feature.

## IV. CONCLUSION AND FUTURE WORK:

The average Wi-Fi signal strength that can be received by the receiver moving throughout the range of access point at any particular speed is found to be nearly constant. As the Wi-Fi signal receiver's speed increases, the peak value of the signal strength received from the source decreases. Hence, the maximum data rate will be less for higher speed. It is possible to predict the mobility of the Wi-Fi signal receiver by knowing the signal strengths received by it within a certain time interval. We can also get insight about greater or lesser speed based on the coefficient of autocorrelation. The complexity of the algorithm in training phase is quadratic, at the same time, complexity involves for speed estimation during testing phase is linear and consequently suitable for real-time applications. Thus, the proposed method is cost-efficient and having low complexity with reasonable accuracy.

Estimation of varying speed of the receiver with limited number of samples (time domain features) in a multi-access points scenario is currently being investigated. The Signal-to-Noise Ratio (SNR) will also be taken into account for the estimation of features and hence, the reliable estimation of varying speed. The parameters of the models may be estimated on the fly for the robust estimation of the mobility in a highly time-varying channel.

## V. ACKNOWLEDGEMENT:

The authors would like to thank Prof. Niki Trigoni and her research group, Oxford University, United Kingdom for granting the permission to use *SensorAp*p for the collection of Wi-Fi samples.